\documentclass[
 reprint,
 amsmath,
 amssymb,
 aps,
 prmaterials,
 floatfix
]{revtex4-2}

\usepackage[utf8]{inputenc}
\usepackage{graphicx} 
\usepackage{dcolumn}
\usepackage[caption=false]{subfig}
\usepackage{bm}
\usepackage[version=4]{mhchem}
\usepackage{xurl}
\usepackage{multirow}
\usepackage{xcolor}
\usepackage[percent]{overpic}
\usepackage{placeins}
\usepackage{graphicx}
\usepackage{subfig} 

\usepackage{graphicx}
\usepackage{subfig}

\usepackage[normalem]{ulem}

\usepackage{hyperref}
\usepackage[nameinlink]{cleveref}

\makeatletter
\renewcommand\p@subfigure{\thefigure}
\makeatother

\crefname{figure}{Fig.}{Figs.}   
\crefname{subfigure}{Fig.}{Figs.}
\crefname{sub@figure}{Fig.}{Figs.}
\crefname{table}{Table}{Tables}
\Crefname{table}{Table}{Tables}
\crefname{equation}{Eq.}{Eqs.}
\Crefname{equation}{Eq.}{Eqs.}

\begin{document}

\title{Competing lattice instabilities in epitaxially strained \texorpdfstring{\ce{BaSnO3}}{BaSnO3}}

\author{Benyao Sun}
\affiliation{ Interdisciplinary Centre for Advanced Materials Simulation (ICAMS), Ruhr-University Bochum, Universitätsstr. 150, 44801 Bochum, Germany
}
\email{benyao.sun@rub.de}
\author{Anna Grünebohm}
\email{anna.gruenebohm@rub.de}
\affiliation{ Interdisciplinary Centre for Advanced Materials Simulation (ICAMS), Ruhr-University Bochum, Universitätsstr. 150, 44801 Bochum, Germany
}
\affiliation{Center for Interface-Dominated High Performance Materials (ZGH), Ruhr-University Bochum, Universitätsstr. 150, 44801 Bochum, Germany}

\begin{abstract}
We use density functional theory to map the lattice‑dynamical landscape of  \ce{BaSnO3} under biaxial compressive strain and uncover the hierarchy of  structural instabilities. 
Our results demonstrate that the phase stability of strained \ce{BaSnO3} cannot be inferred from the softening of the polar mode alone. 
Instead it is governed by the competition among lattice instabilities and their anharmonic coupling to structural relaxation.  
Two‑mode energy‑surface analyses show that the polar and rotational order parameters interact predominantly competitively. 
While $A$- and $M$-point octahedral rotations become unstable before the zone-center polar mode and dominate the harmonic response for moderate values of strain, they are related to a continuous transition without anharmonic lattice coupling and a moderate gain of energy with strain. 
Large strain may induce  a first order transition to a supertetragonal polar state. Once this state is stabilized,  the magnitude of polarization and ferroelectric energy well depth are comparable to \ce{BaTiO3} under compressive strain. 
Furthermore, we discuss how an electric field or electron doping may tip the balance between the different modes and favor of the supertetragonal structure over antipolar and rotational alternatives at a reduced critical strain. 
\end{abstract}

\maketitle

\section{INTRODUCTION}
Barium stannate (\ce{BaSnO3}, BSO) is a thermally and chemically stable cubic perovskite oxide (space group $Pm\bar{3}m$) with a wide band gap of approximately $4$~eV  \cite{Luo2012,Kim2012a}.
Its conduction band is dominated by highly dispersive Sn-$5s$ states associated with a small electron effective mass \cite{Kim2012} and small phonon-scattering \cite{Krishnaswamy2017} allowing for a high room-temperature electron mobility in donor doped BSO. 
The material is therefor relevant as  transparent conductor as well as  for field-effect devices  and high-frequency and high-power oxide electronics \cite{Kim2012a,Luo2012,Park2014,Ngabonziza2024}.
Adding polarization to the material would allow for new tunable functionalities.
However, no structural phase transition has been detected in unstrained BSO down to $10$~K, and first-principles calculations reveal no unstable phonon modes at zero pressure \cite{bevillon2008}.

Over the past few decades, strain engineering  has become the method of choice for tuning the ferroelectric properties of perovskites.
Already percent-level biaxial strains can shift transition temperatures by hundreds kelvin  \cite{Haeni2004,Ederer2005,Schlom2007,Grunebohm2015}, and stabilize ferroelectric polarization in Ti-based incipient ferroelectrics such as \ce{SrTiO3} \cite{Antons2005,Verma2015} and \ce{TiO2} \cite{Grunebohm2011,Deepak2014} as well as in Mn-based perovskites such as \ce{SrMnO3} \cite{Lee2010,An2021}.
This strain-induced or enhanced ferroelectricity along the expanded lattice direction(s) is commonly associated with the softening of polar phonon modes  \cite{Antons2005,Grunebohm2011,Lee2010}.
For large lattice expansion one may eventually drive oxide perovskites toward a  supertetragonal phase with c/a of about 1.3 \cite{Tinte2003,Mellaerts2022}. 
Such strongly strained phases are not restricted to theoretical predictions. 
For example, highly strained T-like or supertetragonal structures have been experimentally stabilized in \ce{BiFeO3} thin films \cite{Kreisel2011}.

So far, it is not established whether the concept of strain-induced supertetragonal and ferroelectric states  can be transferred to BSO with its closed-shell Sn $4d^{10}$ configuration. 
Predictions by density functional theory (DFT) for BSO under compressive epitaxial strain are promising: Zhang \textit{et al.} found a polar ferroelectric state \cite{Zhang2017} and Li \textit{et al.} predicted a first-order transition to a polar supertetragonal phase at approximately $-5\%$  \cite{Li2025}. 
Recently, it was furthermore shown that the lower conduction-band energy of the ferroelectric state may allow to reduce this critical strain by electron doping, e.g.\ by Ba substitution with La
\cite{Zhang2025}.

Although, the predicted value of strain could be realized, e.g., in ultrathin \ce{SrTiO3}/BSO superlattices,  we are however not aware of any experimental realization of ferroelectric BSO. 
On one hand, it is challenging to growth coherent defect lean films of a sufficient thickness with such large mismatch \cite{Prakash2015} and DFT may underestimate the critical strain. 
On the other hand, neither the softening of a polar phonon mode nor the presence of a supertetragonal state in DFT simulations alone are sufficient to guarantee a polar ground state.
Particularly, softening of zone-boundary rotational and antipolar phonon modes \cite{Bousquet2010} may be induced by strain and these may couple either competitively or cooperatively to the polar mode \cite{Sai2000a,Aschauer2014}. 
The coupling may furthermore depend on the magnitude of strain. 
For example, oxygen-octahedron rotations in \ce{SrTiO3} and \ce{CaTiO3} are known to suppress the ferroelectric instability \cite{Eklund2009,Aschauer2014} in the unstrained material, while the coupling in \ce{SrTiO3} changes to cooperative beyond a critical rotation amplitude \cite{Aschauer2014,Gu2018a}. 
Electric fields provide an additional means to control this interplay, as they can directly favor polar distortions and modify octahedral rotations indirectly through their coupling to polarization \cite{Li2019,Hong2013a}.
It is thus timely to reveal the interplay between potential soft phonon modes in \ce{BaSnO3} under the influence of epitaxial strain, electric fields, and electron doping.

Our analysis of the phonon modes and their interaction reveals that zone-boundary rotation modes precede the polar $\Gamma$ instability, while anharmonic coupling to tetragonal strain may induce a first-order transition to a  supertetragonal state that is further favored by an out-of-plane electric field and electron doping.
We estimate an upper bound for the critical strain for the onset of this transition of about $-6.2\%$, and show how competing  $A$-point octahedral rotations may further increase the critical strain.

\section{COMPUTATIONAL DETAILS}
First-principles calculations were performed within density functional theory (DFT) \cite{Kohn1965}, as implemented in the ABINIT package \cite{Verstraete2025}, using the Perdew--Wang 1992 (PW92) local density approximation (LDA) exchange-correlation functional \cite{Perdew1992}.
The projector augmented-wave (PAW) method \cite{Torrent2008} was employed, using PAW data sets from the Jollet--Torrent--Holzwarth (JTH) table \cite{Jollet2014}.
The $5\mathrm{s}^{2}5\mathrm{p}^{6}6\mathrm{s}^{2}$ electrons of \ce{Ba},
$4\mathrm{d}^{10}5\mathrm{s}^{2}5\mathrm{p}^{2}$ electrons of \ce{Sn}, and
$2\mathrm{s}^{2}2\mathrm{p}^{4}$ electrons of \ce{O} were treated as valence electrons.
Relative energies were converged to  1~meV/atom with a plane-wave kinetic-energy cutoff of 70~Ha together with a $6\times6\times6$ Monkhorst--Pack $k$-points mesh. 
The threshold for electronic self-consistency was set to either a total-energy difference between successive electronic iterations below $10^{-12}$ Ha or the variation in the maximum force less than $2\%$.
Structural relaxation was performed until the residual forces on all atoms were below $10^{-6}$~Ha/Bohr ($\approx 5\times10^{-5}$~eV/\AA). 
The spontaneous polarization was computed using the Berry-phase approach \cite{Berryphase1993}.

To model coherent biaxial epitaxial strain in the (001)-plane, we fixed the in-plane lattice constants ($a = b$) and the cell angles, while the out-of-plane lattice constant $c$ was fully relaxed. 
Epitaxial strain $\eta$ of  $-1\%$ to $-9\%$ ($\Delta \eta = $1\%) was sampled relative to  $a_0=4.088$~\AA, the equilibrium lattice constant,  as:
\begin{equation}
\eta=\frac{a-a_{0}}{a_{0}}\times 100\,\%\;.
\label{eq:strain}
\end{equation}
For each strain, the phonon dispersion along the high-symmetry path of the first Brillouin zone was calculated on a regular $6\times6\times6$ $q$-point grid and  Fourier-interpolated along the high-symmetry directions 
using density-functional perturbation theory (DFPT) \cite{Gonze1997b}. 
During phonon post-processing, the  acoustic sum rule was imposed on the interatomic force constants and LO--TO splitting at $\Gamma$ was added by  non-analytic correction, using the DFPT-calculated high-frequency dielectric tensor $\epsilon^\infty$ and Born effective charges $Z_i^*$ \cite{Gonze1997}.

To record the energy landscapes beyond the harmonic approximation, first, the phonon eigenvectors $\bm{u}$ and $\bm{w}$  scaled with scalar factors $\lambda'$ and $\alpha$ were frozen-in as 
\begin{equation}
  \bm{S}(\lambda')
  =
  \bm{S}_{0}
  +
  \lambda'\bm{u} + \alpha \bm{w},
\label{eq:interpolationdis}
\end{equation}
where $\bm{S}_{0}$ is the high symmetry structure  (positions and lattice parameters). 
Thereby, the c-lattice parametes were relaxed. Second, for all  obtained local  energy minima, a full relaxation of 
atomic positions were done.
While the atomic relaxation is negligible for the rotational modes, the combined relaxation of atomic positions and c-lattice parameter yields a supertetragonal state in case of the  $\Gamma$-mode. 
To go beyond the frozen $\Gamma$-mode landscape, complementary the energy landscape was then obtained by interpolating between the high-symmetry reference and this relaxed supertetragonal structure according to 
\begin{equation}
  \bm{S}(\lambda)
  =
  \bm{S}_{0}
  +
  \lambda
  \left(
  \bm{S}_{\mathrm{ST}}
  -
  \bm{S}_{0}
  \right),
\label{eq:interpolation}
\end{equation}
where $\bm{S}_{\mathrm{ST}}$
denotes the structure of the supertetragonal structure.

For clarity, we compare $S(\lambda)$ and $S(\lambda')$  not by their displacement amplitudes but rather by their polarization along $z$ approximated as 
\begin{equation}
\vec{P}_z\approx \frac{e}{V} \sum_i \,, Z_{i,zz}^*\vec{u}_{i,z} 
\end{equation}
with $\vec{u}_{i,z}$ being the displacement of the $i$th atom from its position in the high-symmetric reference structure and $V$ is the unit-cell volume. 

To record the polar off-centering  in a larger strain range, complementary, full relaxation of atomic $z$ coordinates and the out-of-plane lattice parameter has been performed after imposing a small displacement along the polar $\Gamma$-point mode eigenvector in a single unit cell. 

\section{RESULTS}

\begin{figure}[tpb]
  \centering
  \includegraphics[width=\linewidth]{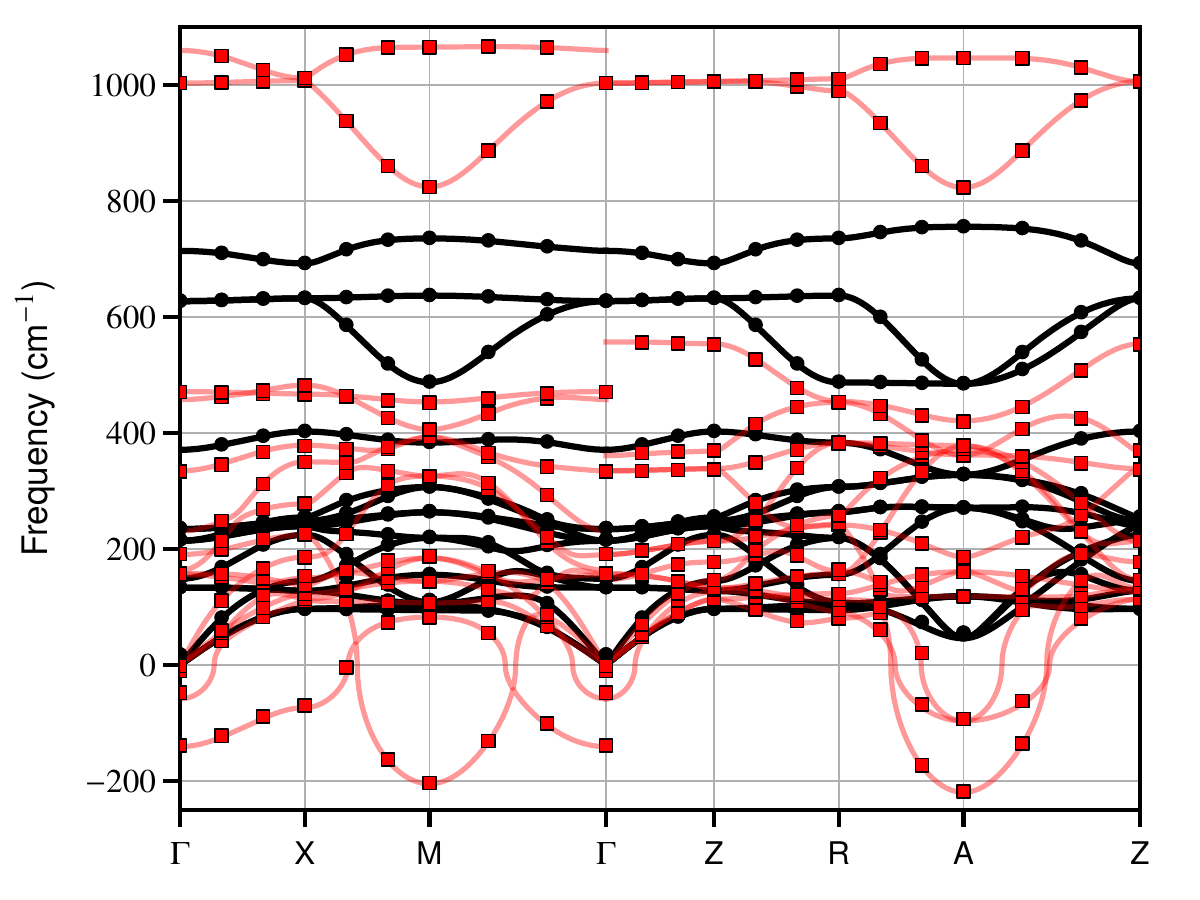}
  \caption{Phonon dispersions of BSO in the cubic phase (black) and in the high-symmetry tetragonal phase under $-8\%$ strain (red).
  For comparison, both dispersions are shown along the same reduced-coordinate $q$-path and labeled according to the tetragonal Brillouin zone.
  Symbols and lines denote explicitly calculated frequencies and interpolated phonon branches, respectively.
  Imaginary frequencies, are shown as negative values.}
  \label{fig:phonon}
\end{figure}

\begin{figure}[tbp]
  \centering
  \includegraphics[width=.8\linewidth]{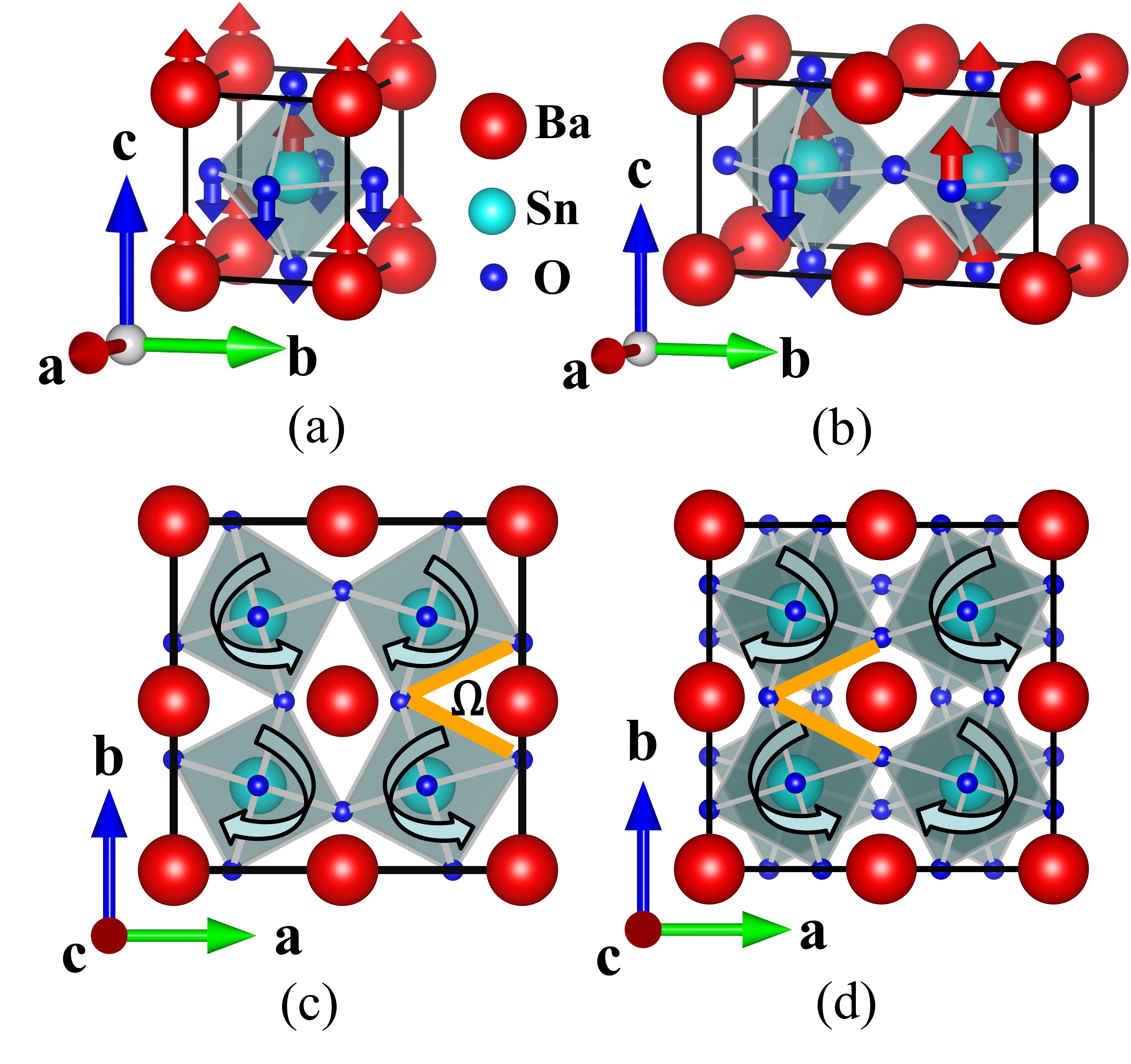}%
  \caption{Displacement patterns of the 
   phonon modes in BSO softening under compressive in-plane strain.
  (a) $\Gamma$-point: polar distortion.
  (b) $X$-point: antipolar distortion.  
  (c) $M$-point: in-phase octahedral rotation ($a^{0}a^{0}c^{+}$).
  (d) $A$-point: out-of-phase octahedral rotation ($a^{0}a^{0}c^{-}$).
  Arrows denote atomic displacements in (a, b) and rotations of the octahedra about the $c$ axis in (c, d).
  The rotation angle is measured as $\theta$ =  $(90^\circ-\Omega)/2$, where $\Omega$ denotes the O--O--O angle (orange lines) formed by three oxygen ions belonging to two neighboring coner-sharing octahedra in the $x-y$ plane. } 
  \label{fig:softmodes_illustration}
\end{figure}

Without biaxial strain, the equilibrium lattice constant $a_0$ calculated using the LDA pseudopotential is $a_0=4.088$~\AA, which is close to the previously reported theoretical value of 4.070~\AA  \cite{Monserrat2018} and slightly smaller than the experimental value of 4.116~\AA  \cite{maekawa2006}. 
For this lattice constant, the cubic phase of BSO is dynamically stable and shows no imaginary phonon frequencies in agreement with previous first-principles studies \cite{bevillon2008} (Black curves in \cref{fig:phonon}). 
For the chosen technical details, the lowest polar optical mode at $\Gamma$ lies at approximately $140~\mathrm{cm^{-1}}$ and the degenerate modes at $A$ and $M$ point are slightly below.

Under compressive strain, the c-lattice parameter increases and for large strain, this tetragonal reference structure becomes unstable with respect to phonon modes  (Red curves in \cref{fig:phonon} for 8~\%).
First, the lowest mode around the $\Gamma$-point corresponds to the polar   shift of cations vs.\ anions  along the out-of-plane [001] direction (\cref{fig:softmodes_illustration}(a)).
A second, weaker instability at $\Gamma$ (not shown) corresponds to nonpolar oxygen buckling, in which the two equatorial oxygen atoms move oppositely along [001].
The polar instability is not confined to $\Gamma$ but spans the whole $\Gamma$--$X$ path and approximately the first half of $\Gamma$--$M$. This means the reference phase is not stable with respect to antipolar displacements of different periodicity (\cref{fig:softmodes_illustration}(b) for $X$), similar to the polar instability of cubic \ce{BaTiO3} \cite{Ghosez1998}.
Note that only polar displacements along [001] are low in energy. 
Due to the LO-TO splitting, there is thus no (anti)polar soft mode branch along $Z$ or $R$.

\begin{figure}[tbp]
  \centering
  \includegraphics[width=0.55\linewidth]{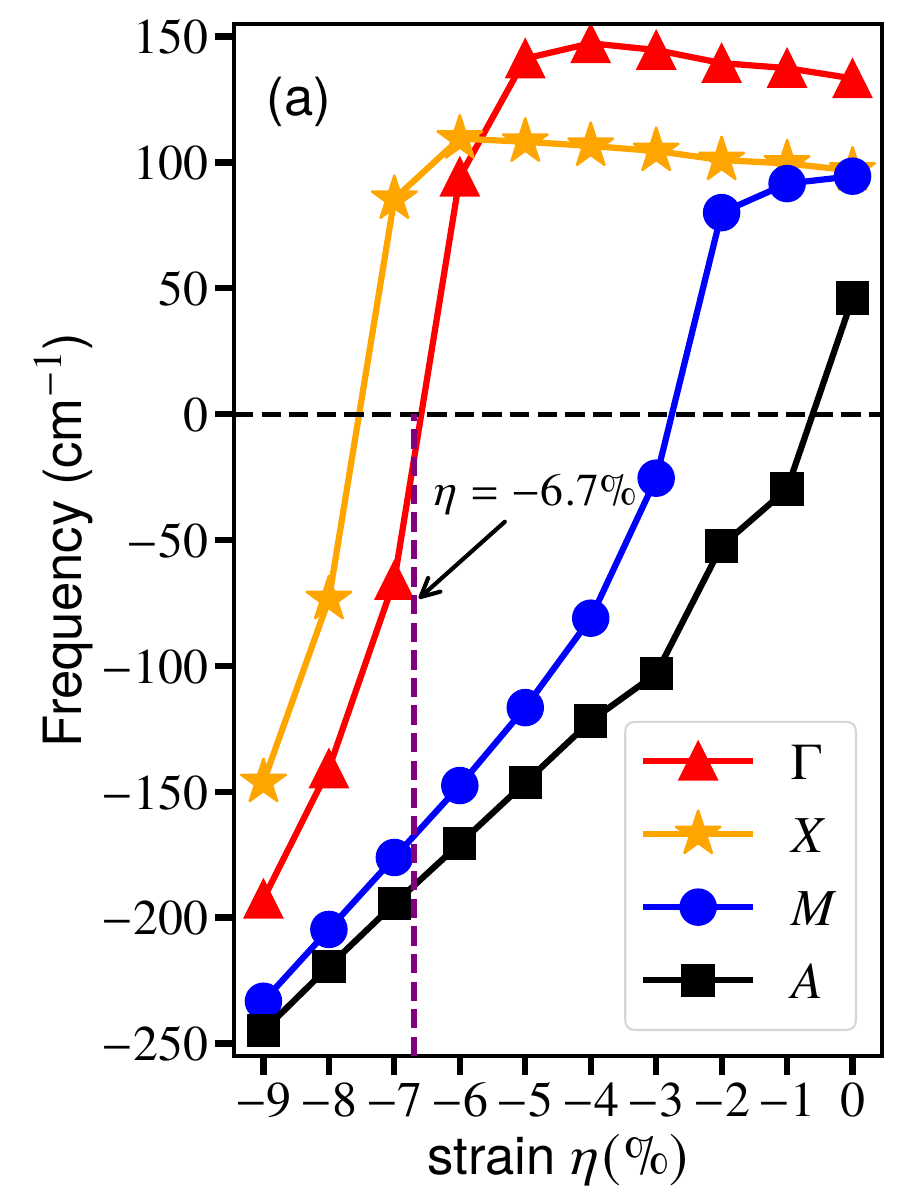}
  \caption{Frequencies of the lowest modes at the $\Gamma$, and $X$, $A$, $M$ points (see \cref{fig:softmodes_illustration}) as a function of compressive strain. 
  Imaginary frequencies, are shown as negative values. 
  Note that dense sampling (0.1\%) has been used only for the $\Gamma$-point mode, with the corresponding values given in \cref{tab:strain_frequency} of the Appendix. 
} 
  \label{fig:minimum_frequ_per_mode}
\end{figure}

The strongest harmonic instabilities, however, are not polar.
The largest imaginary frequencies are related to rotations of the \ce{SnO6} octahedra about the tetragonal $c$ axis either in-phase ($a^{0}a^{0}c^{+}$, \cref{fig:softmodes_illustration}(c)) or out-of-phase ($a^{0}a^{0}c^{-}$, \cref{fig:softmodes_illustration}(d)) around the $M$ and $A$ point, respectively.
Furthermore the high symmetry structure is not stable against  out-of-phase rotations around $x$ and $y$ ($a^{-}b^{0}b^{0}$ and $a^{0}b^{-}c^{0}$, not shown) at the $A$ point for $-8\%$. 
This degenerate modes are however not soft for smaller values of strain and higher in energy and thus neglected in the following.

\cref{fig:minimum_frequ_per_mode} compares the strain dependence of the lowest-frequency phonon modes at each high-symmetric point.
The lowest $A$-point rotation becomes unstable, already at approximately $-1\%$, followed by the $M$-point rotation near $-3\%$, while the polar and antipolar modes follow at about $-6.7\%$, and around $-8\%$.
Although the change of the frequency with strain beyond these critical values of strain is largest for the polar mode, the octahedra rotation modes has the  larger imaginary frequencies even up to $-9\%$ strain. 

\begin{figure}[tbp]
  \vspace*{-3.5mm}
  \centering
  \subfloat{%
    \includegraphics[height=0.65\linewidth,clip,trim=0cm 0cm 0cm 0cm]{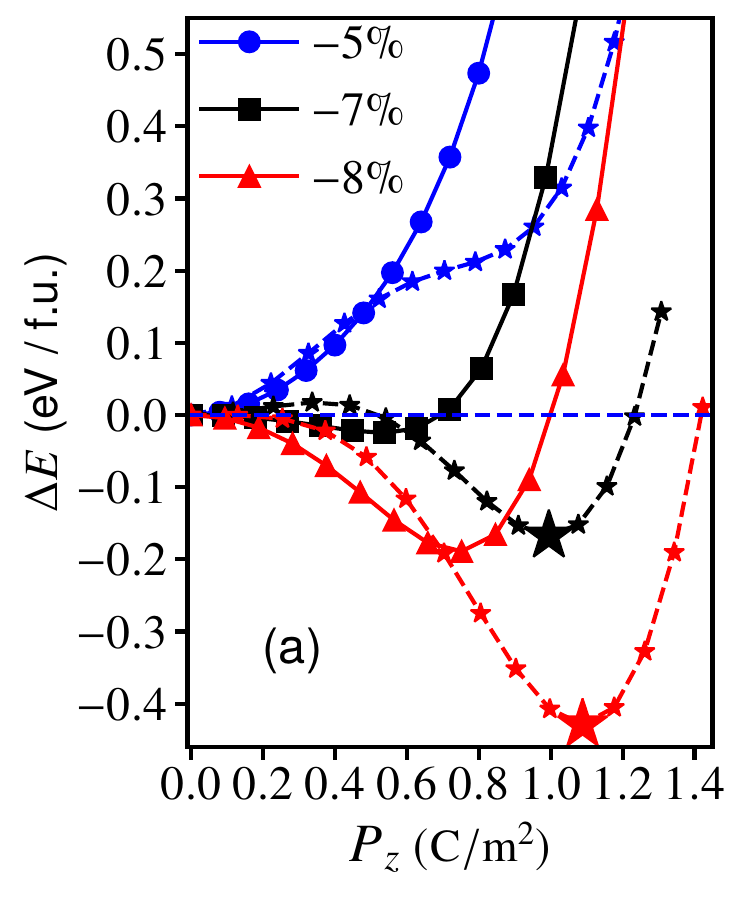}%
    \label{fig:Gamma_energy}%
  }
  \subfloat{%
    \includegraphics[height=0.65\linewidth,clip,trim=3cm 0cm 0cm 0cm]{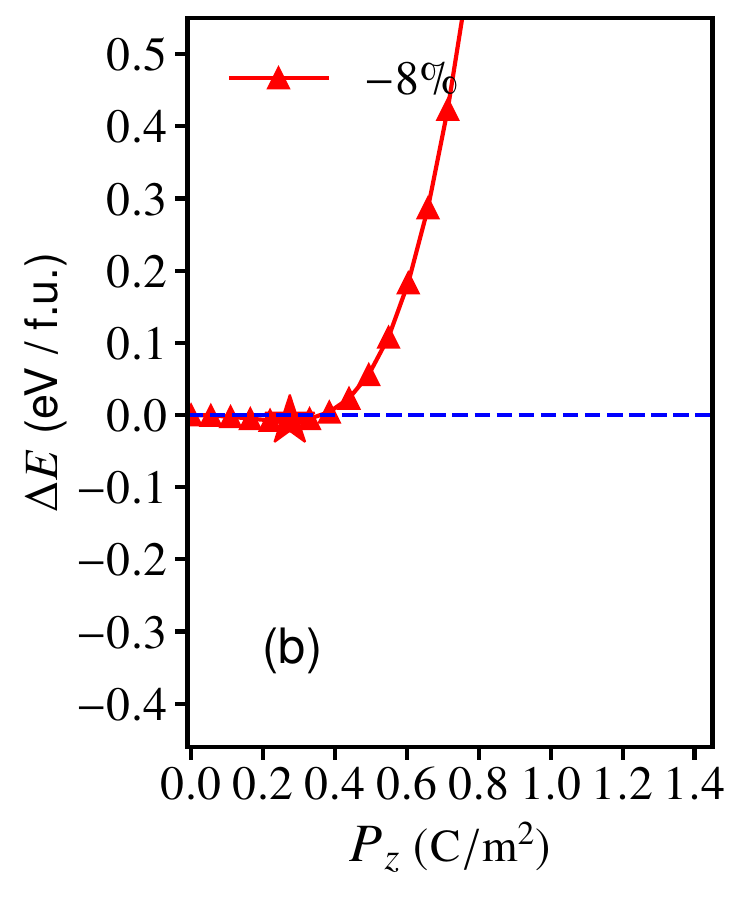}%
    \label{fig:X_energy}%
  }
  
  \vspace{-1.5em}
  
  \subfloat{%
    \includegraphics[height=0.65\linewidth,clip,trim=0cm 0cm 0cm 0cm]{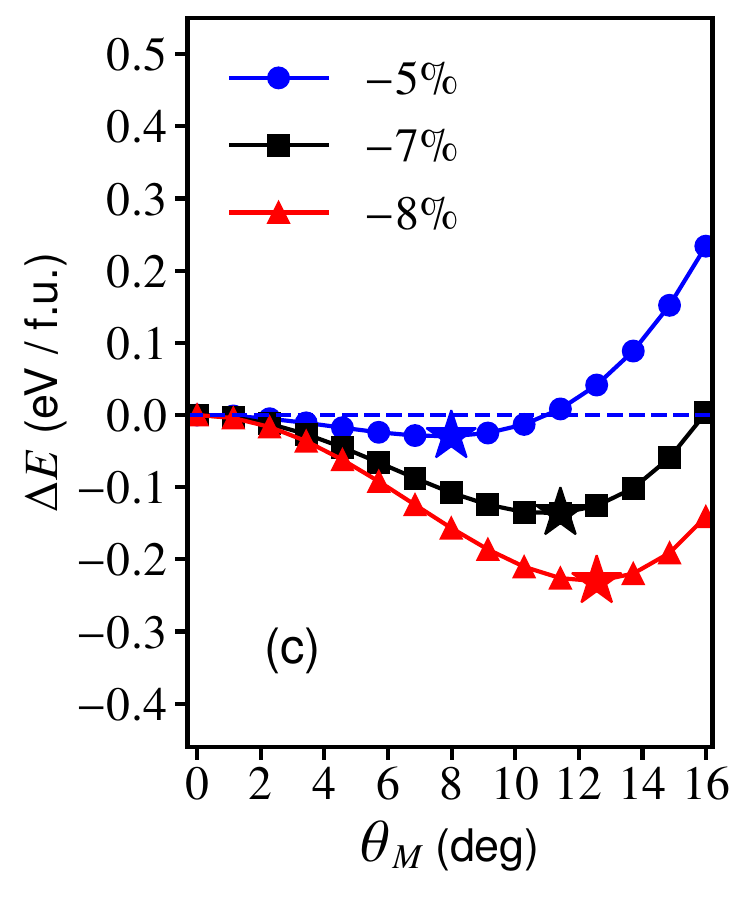}%
    \label{fig:M_energy}%
  }  
  \subfloat{%
    \includegraphics[height=0.65\linewidth,clip,trim=3cm 0cm 0cm 0cm]{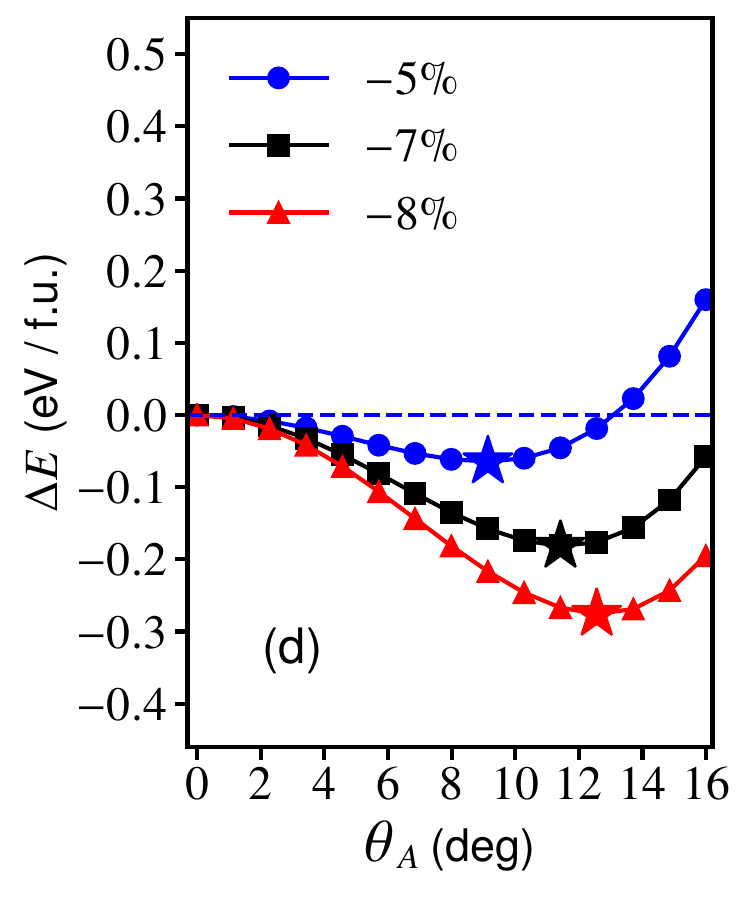}%
    \label{fig:A_energy}%
  }
  \caption{Energy landscapes obtained by freezing in the 
  (a) $\Gamma$, 
  (b) $X$,
  (c) $M$, and 
  (d) $A$-point phonon modes under compressive strain (solid lines, Eq.\eqref{eq:interpolationdis} with $\bm{w}=0$).
  For the $\Gamma$- and $X$-point modes, $P_z$ denotes the mode polarity (the local primitive-cell polarity for the antipolar $X$-point mode), whereas $\theta$ denotes the octahedral rotation angle for the $M$- and $A$-point modes.
  Large stars indicate structures obtained after further relaxation of the frozen-mode minima at fixed in-plane lattice constants.
  The dashed curves with stars in (a) show the energy along a linear interpolation between the centrosymmetric reference and the relaxed displacement pattern (Eq. \eqref{eq:interpolation}).}  
  \label{fig:E_GammaXMA}
\end{figure}

To crosscheck if the harmonic picture is sufficient at such large compression, we record the energy landscape for the frozen-in eigenvector with relaxed $c$-lattice parameter  separately for every soft mode  (solid lines in \cref{fig:E_GammaXMA}). 
Indeed shallow double-wells build up for the polar $\Gamma$-point mode and the antipolar $X$-point only above  $-6.7\%$  (\cref{fig:E_GammaXMA}(a)) and above $-7\%$  (\cref{fig:E_GammaXMA}(b)) strain, while pronounced minima are visible for  $M$-point (\cref{fig:E_GammaXMA}(c)) and $A$-point (\cref{fig:E_GammaXMA}(d)) already for the smallest shown strain of $\eta=-5\%$.
With increasing compression, all wells deepen and the minima shift to larger values of polarization or rotation angle,  reaching about $12.6^\circ$ at $\eta=-8\%$. 

Full relaxation of atomic positions and c-lattice starting from these energy  minima,  does neither change atomic configuration nor energy for $X$, $A$ and $M$ modes. 
Notably, relaxing the $\Gamma$-point related minima results in a new ground state structure (stars in \cref{fig:E_GammaXMA}(a)) with larger $P_z$ and lower energy.
For $-7\%$, the energy goes down  by $0.142~$eV/f.u.\  and the polarization reaches about 1~C/m$^2$. In this process, the relative Ba--O  displacements become comparable to the Sn--O displacement which dominats in the harmonic $\Gamma$-mode eigenvector. 
As this relaxation is also related to a drastic increase of the c/a ratio from approximately $1.12$ to above $1.35$, we follow the nomenclature from \cite{Kreisel2011} and call this state supertetragonal in the following.

\begin{figure}[tbp]
  \centering
  \includegraphics[width=0.6\linewidth]{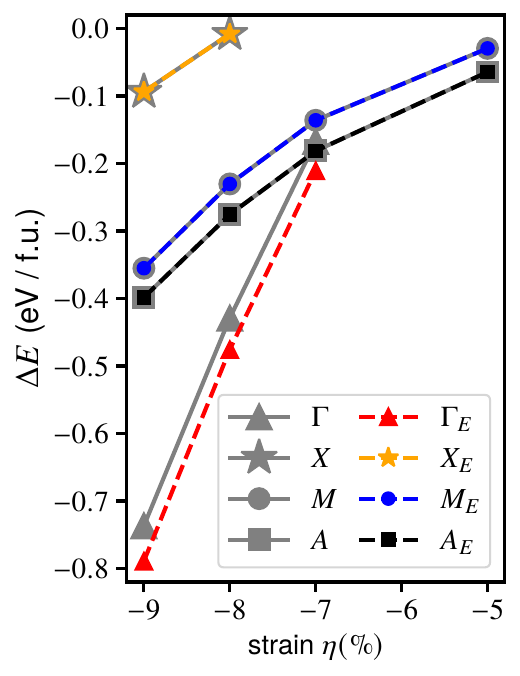}
  \caption{Change of the energy gain by polar and antipolar distortions (triangles, stars) and M and A-point rotations (circles, squares) with strain after full relaxation of atomic positions and c-lattice parameter.
  The gray curves show the energy minima extracted from \cref{fig:E_GammaXMA}, whereas the colored curves show the corresponding minima under an electric field of $E_z=+1028$~kV/cm.} 
  \label{fig:eeta}
\end{figure}

Once in the supertetragonal state, the energy difference to the reference state increases linearly with $|\eta|$ and the polar mode is lower in energy than the $X$, $M$ and $A$-point modes (\cref{fig:eeta}). 
This also provides a qualitative indication of the phase transition temperature associated with the supertetragonal state. 
The energy difference of $0.17$~eV/f.u.\ under $-7\%$ strain is exceeding reported cubic--tetragonal energy differences of approximately $0.01$~eV/f.u.\ for \ce{BaTiO3} and $0.06$~eV/f.u.\ for \ce{PbTiO3} \cite{Yuk2017}, with experimental transition temperatures of approximately $393$~K and $766$~K, respectively \cite{Kirbus2024,Zelezny2015}.
Furthermore, the polar supertetragonal state can be further stabilized an electric field along $c$.
For example, for $E_c=1028~\mathrm{kV/cm}$ the energy of the supertetragonal state is lowered by about $-0.045~\mathrm{eV/f.u.}$ for all values of strain, whereas the $X$-, $M$-, and $A$-derived states do not directly couple to an electrical field. 

Dashed lines in \cref{fig:E_GammaXMA}(a) show the energy landscape for the linear interpolation between the centrosymmetric and supertetragonal states  [\cref{eq:interpolation}].
For $\eta=-7$\%, a deep double well potential around the relaxed minimum exists which is separated from the non-polar state by an energy barrier of 0.017~eV. 
This is a first indication of the first order character of the transition to the polar phase. This interpretation is further supported by the energy landscape for $\eta=-5\%$ using $\bm{S}(\lambda)$ as found for $-7\%$ (blue dashed line). 
This interpolation pattern  shows an incipient instability at $0.7~\mathrm{C/m^2}$. 

\begin{figure}[tbp]
  \centering

  \subfloat{%
    \includegraphics[width=.9\linewidth]{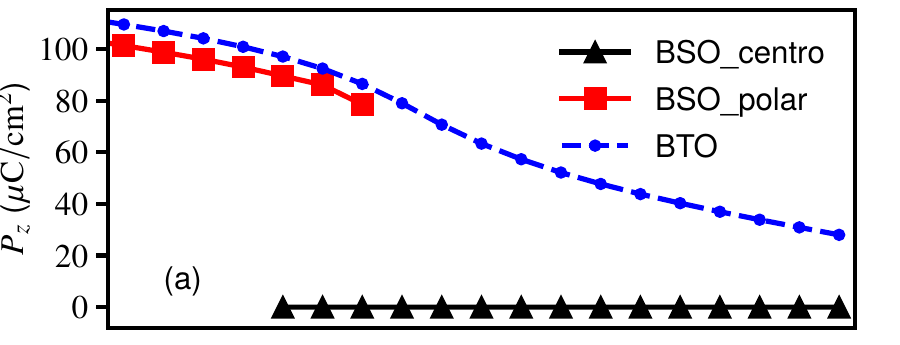}%
    \label{fig:Polarization}%
  }
  
  \par\vspace{-14pt}

  \subfloat{%
    \includegraphics[width=.9\linewidth]{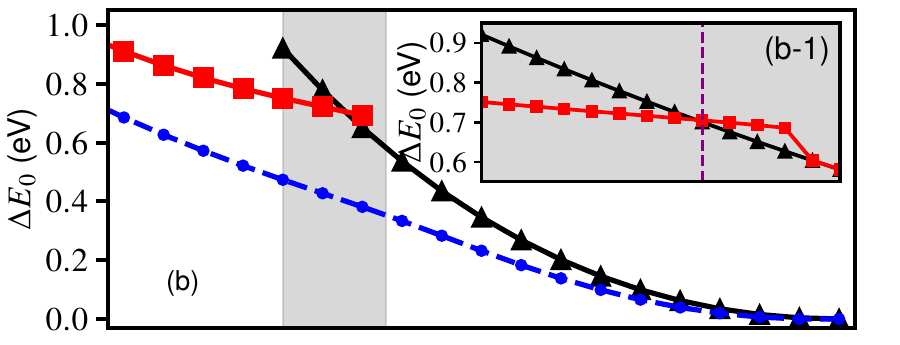}%
    \label{fig:energy}%
  }
  
  \par\vspace{-14pt}

  \subfloat{%
    \includegraphics[width=.9\linewidth]{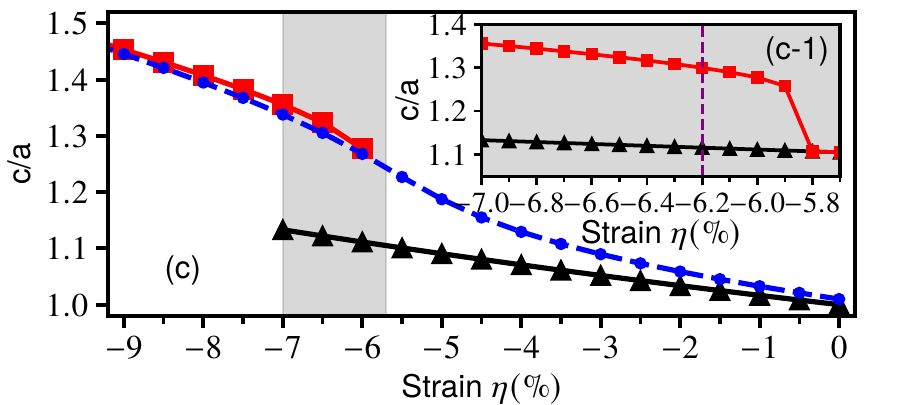}%
    \label{fig:ratio}%
  }
  
  \subfloat{%
    \includegraphics[width=.85\linewidth]{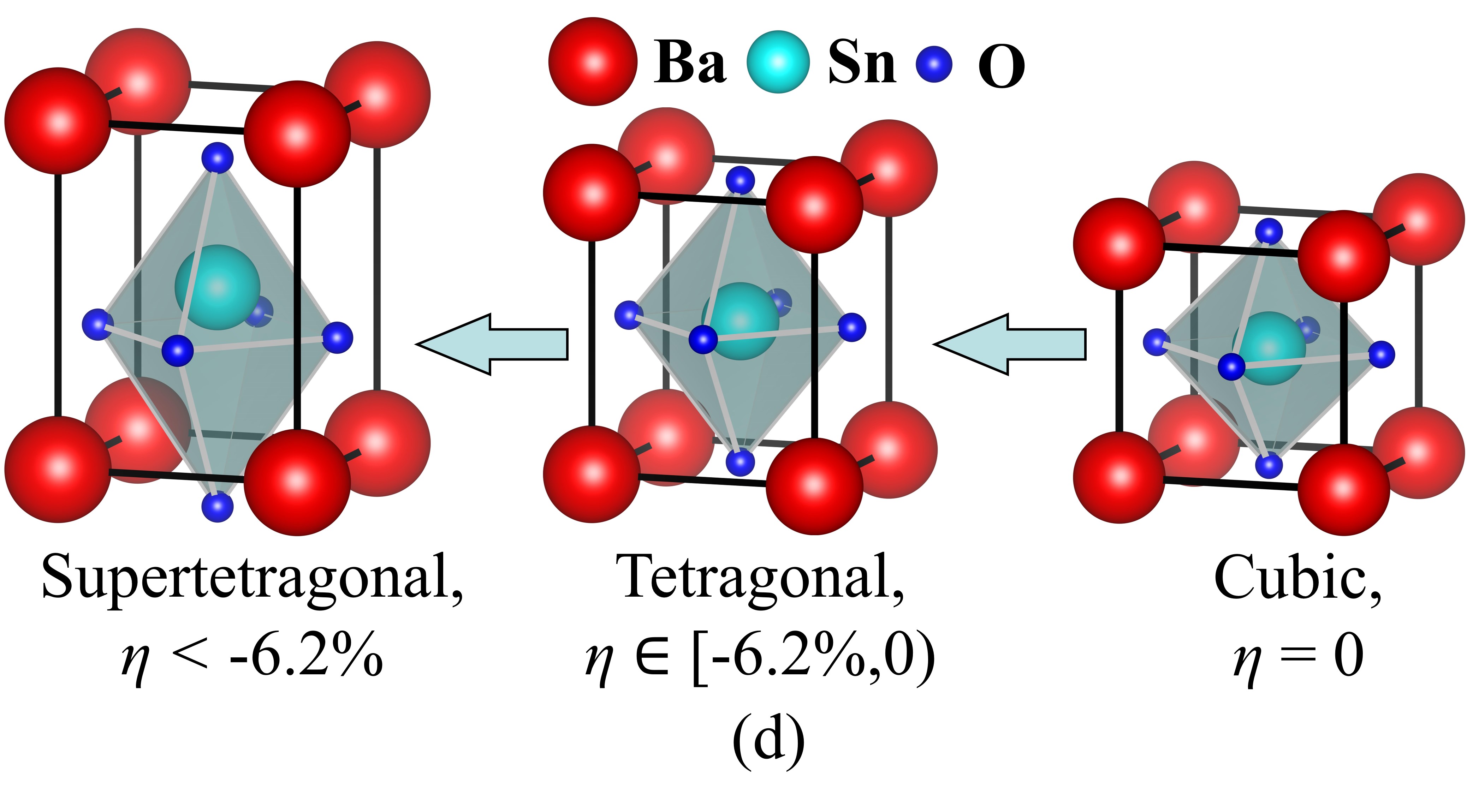}%
    \label{fig:cubic_to_tetragonal}%
  }
  \caption{Comparison of the evolution of, 
  (a) Polarization $P_z$, 
  (b) energy difference $\Delta E_{0}$ relative to the high-symmetry structure without strain and (c) tetragonal ratio $c/a$, with compressive strain $\eta$ between BSO and tetragonal BTO (blue) without octahedral rotations.  
  Insets (b-1) and (c-1) show the  gray-shaded strain range with larger resolution.
  For $\eta<-6.2$~\%, the supertetragonal polar phase (red) is lower in energy than the centrosymmetric one (black). 
  (d) Schematic evolution of BSO from cubic to tetragonal and supertetragonal phases with increasing compression.
}
  \label{fig:ratioall}
\end{figure}

The pronounced relaxation of the $\Gamma$-mode minimum and the first-order character of the energy landscape  raises the question of the strain range of a (meta)-stable supertetragonal state. 
As the used relaxation procedure does not  lower the symmetry, the centrosymmetric state can  be stabilized for arbitary values of strain. 
For convenience we only sample it to $-7$\%. 
In this state, the polarization is zero by definition while $c/a$ increases continuously with compression and reaches approximately $1.1$ near $-6.2\%$ (black lines in \cref{fig:ratioall}). 

While, linear interpolation of the harmonic phonon frequencies yields a critical strain of $-6.7$~\%, the ferroelectric supertetragonal state is already the energetic ground state for $\eta <-6.2\%$
and is metastable up to $\eta <-5.9\%$ (see \cref{fig:ratioall}(b). 
Note that previous work based on GGA functionals found a smaller critical strain of $-5\%$ \cite{Li2025} relative to a lattice constant about $2\%$ larger than the experimental one. 
Our LDA simulations underestimating the lattice constant by about $1\%$ thus give an upper bound for the critical strain.

Once in the supertetragonal state, the polarization abruptly reaches  $90$~$\mu\mathrm{C\,cm^{-2}}$. 
This large value of polarization is however not different to the polarization of \ce{BaTiO3}  under the same value of strain (blue lines in \cref{fig:ratioall}).
Increasing $|\eta|$ further, induces a smooth change of $P_z$ and $c/a$ as in tetragonal \ce{BaTiO3} in agreement to literature \cite{Choi2004a,Neaton2011a}.
The key difference between BSO and prototype ferroelectrics is thus neither the magnitude of polarization nor of the tetragonal strain, but it is the abrupt onset of the ferroelectric phase at a large value of strain.

\begin{figure}[tbp]
  \centering
  \subfloat{%
    \includegraphics[width=0.85
    \linewidth]{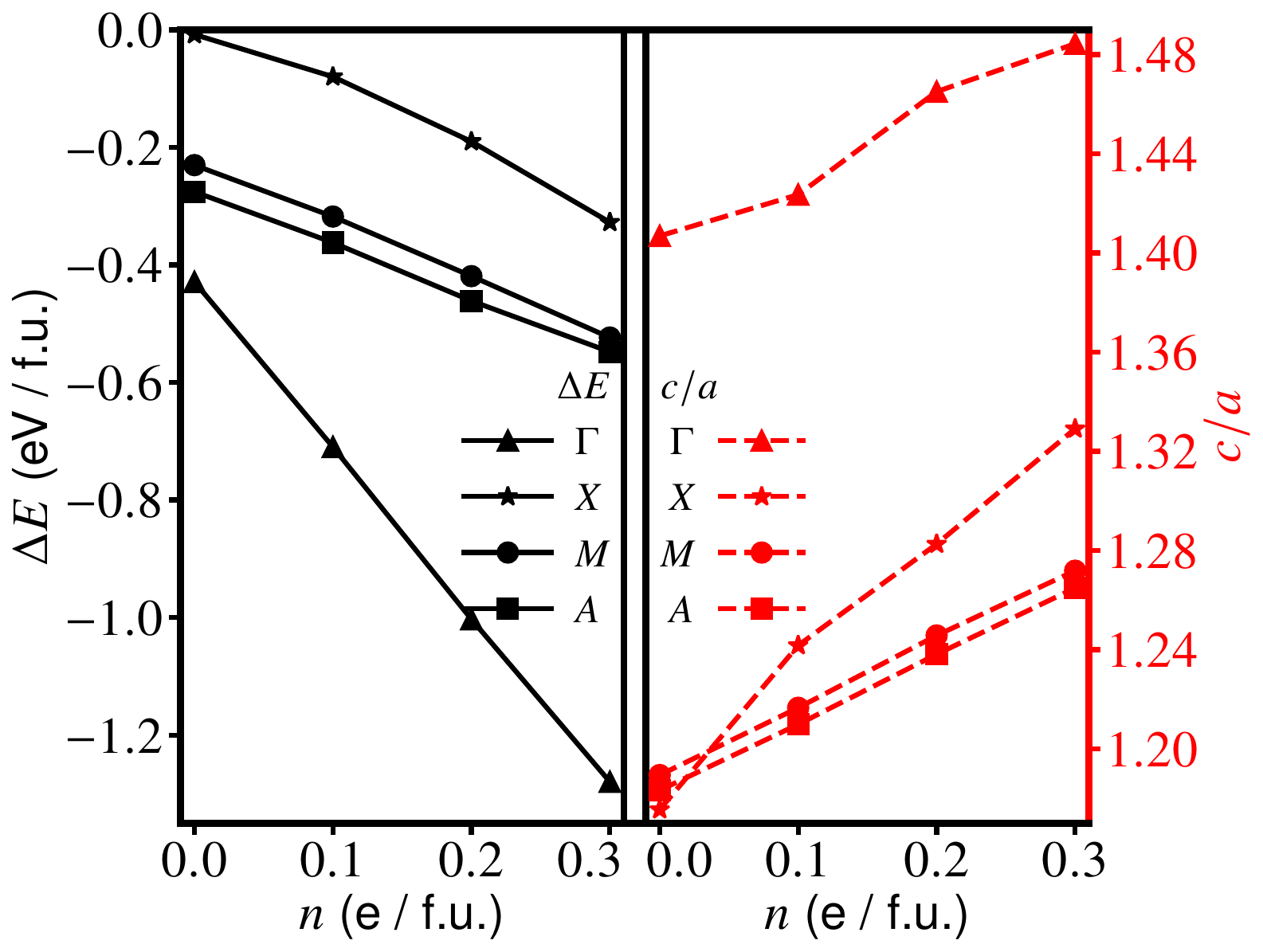}%
  }

  \caption{Change of energy gain (left, black axes) and tetragonal ratio (right, red axes) by polar and antipolar distortions (triangles, stars) and M and A-point rotations (circles, squares) distortions with electron doping for BSO and $\eta=-8\%$.
  }
  \label{fig:electron_doping}
\end{figure}

One possibility to reduce the critical strain for the onset of ferroelectricity is electron doping. For $\eta=-8\%$, $n=0.3~e/\mathrm{f.u.}$ increases the energy difference between non-polar and supertetragonal state from  $-0.42~\mathrm{eV/f.u.}$ to  $-1.28~\mathrm{eV/f.u.}$ (line with black triangles in \cref{fig:electron_doping}).
It is interesting to note, that all four types of instablities soften further with electron doping.
However, the stabilization associated with the octahedral rotation modes at the $M$ and $A$ points is much weaker, with the energies decreasing by only approximately $0.27~\mathrm{eV/f.u.}$ at this doping level.
Both the softening of the modes and its differences between modes is attributed to the conduction band, which is lower in the distorted structures \cite{Zhang2025}. While this offset is about $1.5~\mathrm{eV}$ in the supertetragonal structure it is $\sim 0.5~\mathrm{eV}$ only for the rotation modes (see \cref{fig:dos} in the Appendix).\\

\begin{figure}[tbp]
  \centering

  \subfloat{%
    \includegraphics[
      width=0.468\linewidth,
      clip,
      trim=0cm 2.3cm 4cm 0cm
    ]{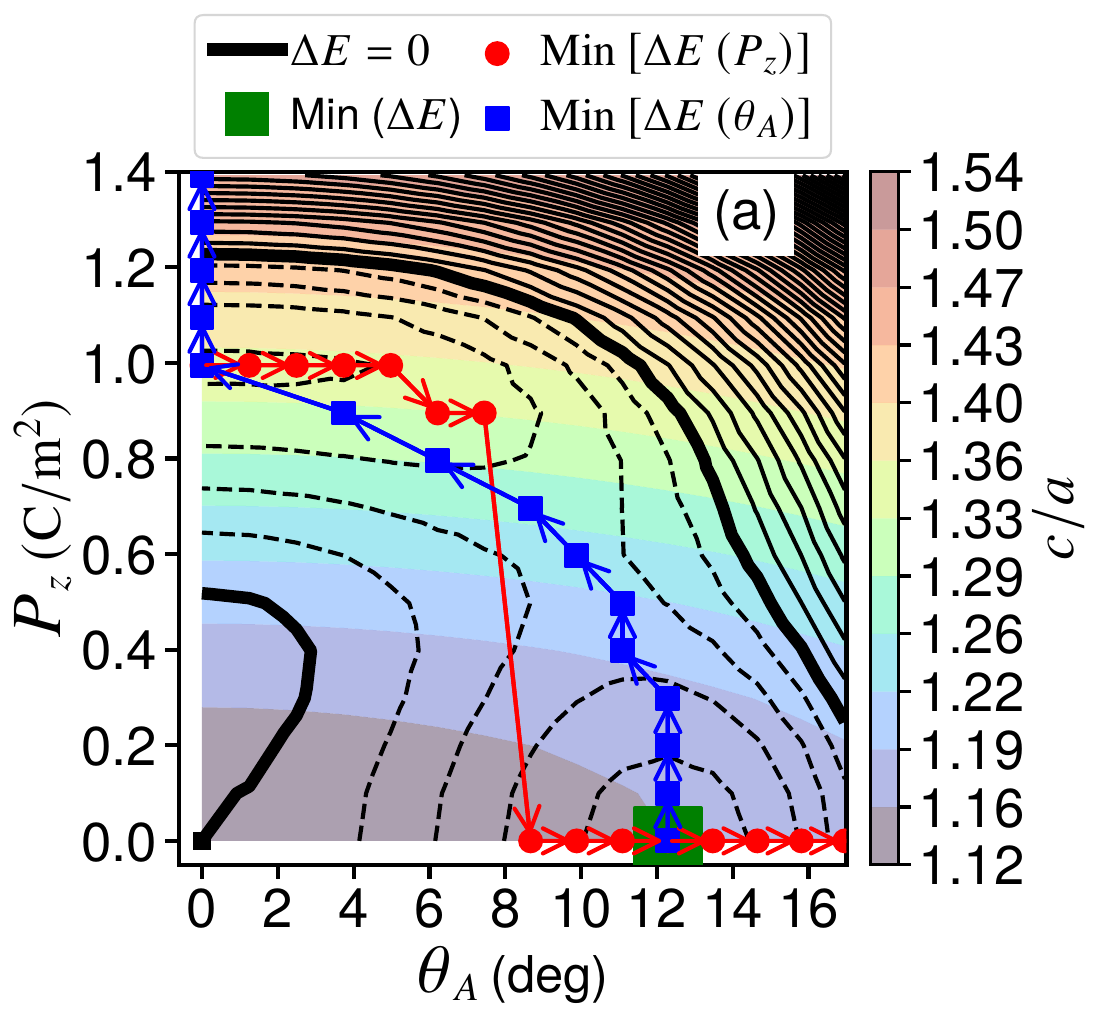}%
    \label{fig:GAcomp7}%
  }
  \hspace{-0.2em}
  \subfloat{%
    \includegraphics[
      width=0.498\linewidth,
      clip,
      trim=3cm 2.3cm 0cm 0cm
    ]{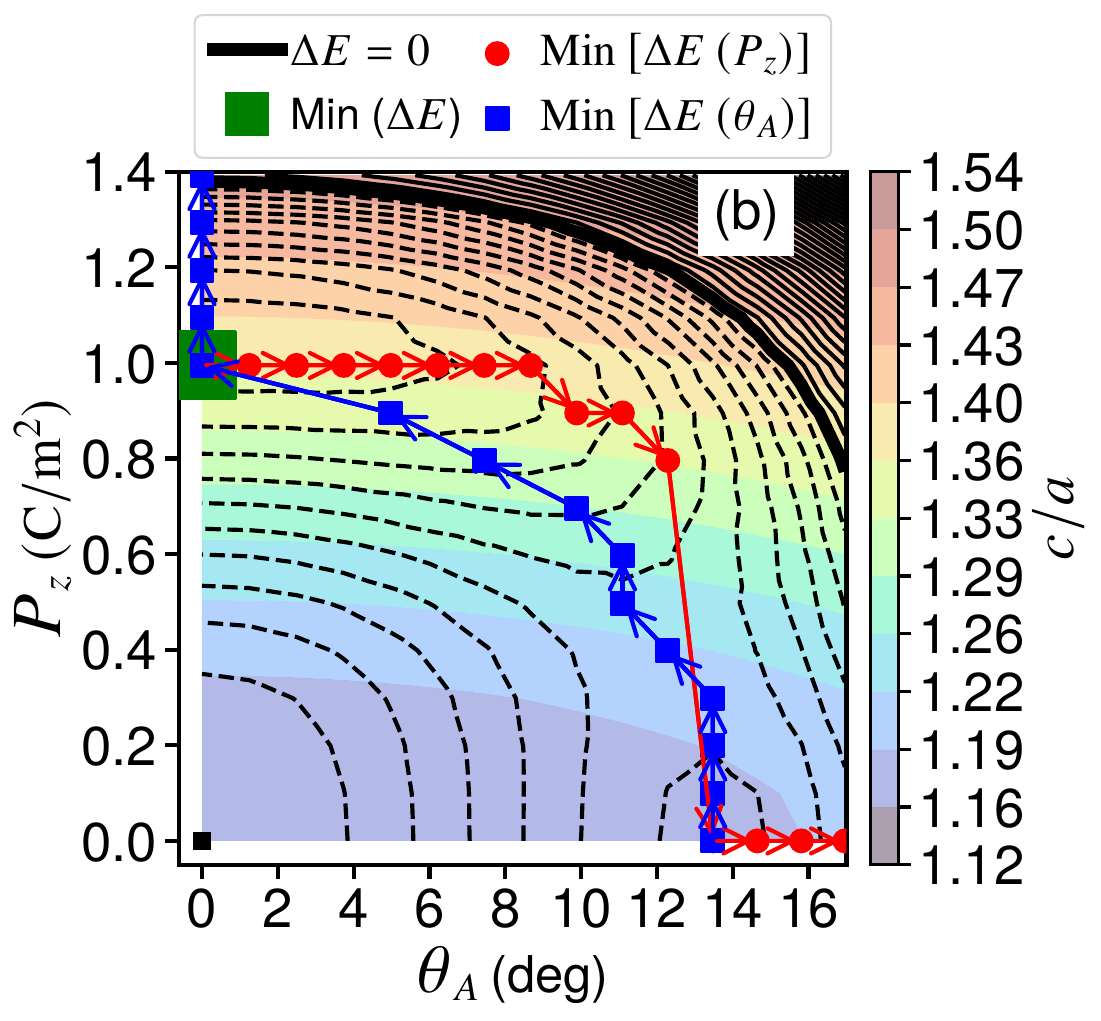}%
    \label{fig:GAcomp7e01}%
  }

  \vspace{-1em}

  \subfloat{%
    \includegraphics[
      width=0.468\linewidth,
      clip,
      trim=0cm 0cm 4cm 2.5cm
    ]{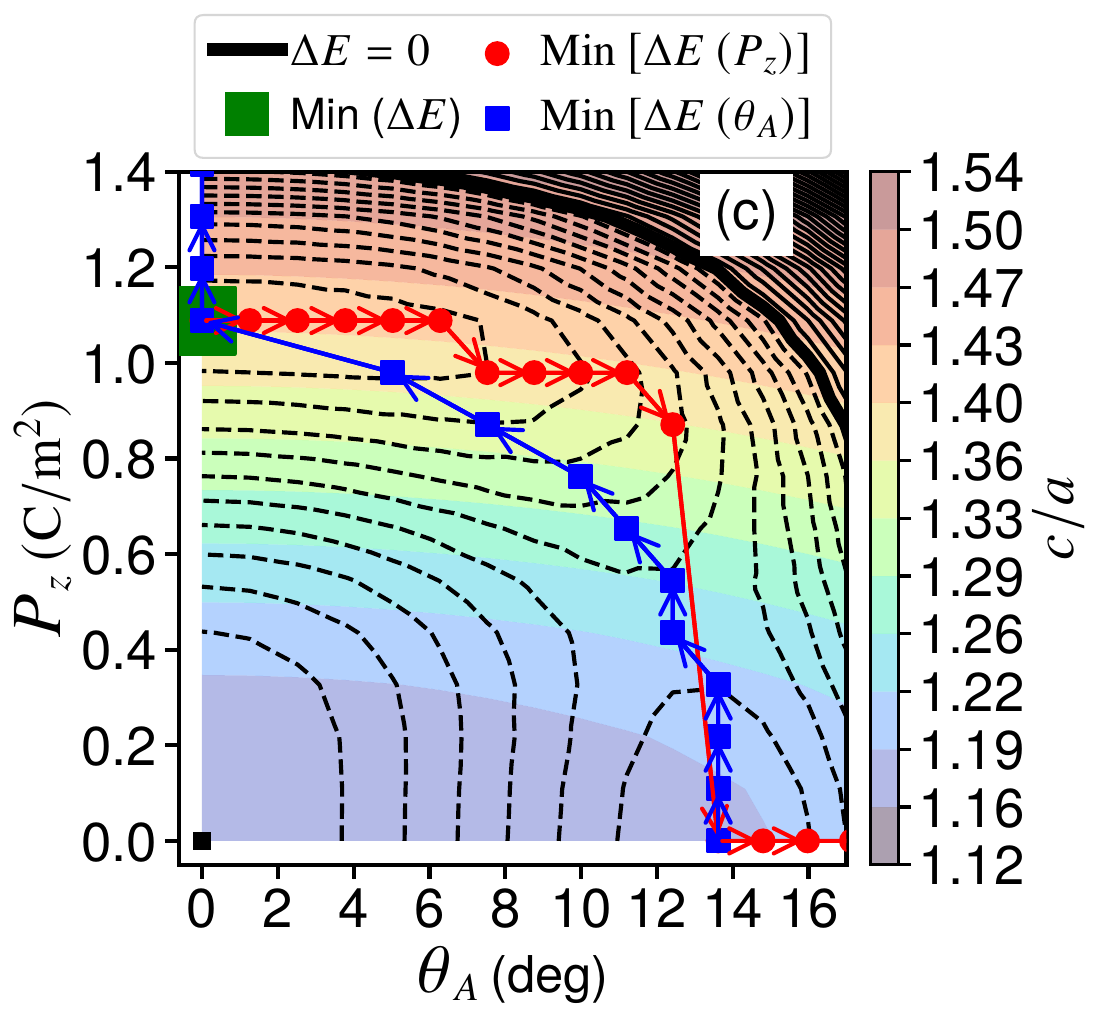}%
    \label{fig:GAcomp8}%
  }
  \hspace{-0.2em}
  \subfloat{%
    \includegraphics[
      width=0.498\linewidth,
      clip,
      trim=3cm 0cm 0cm 2.5cm
    ]{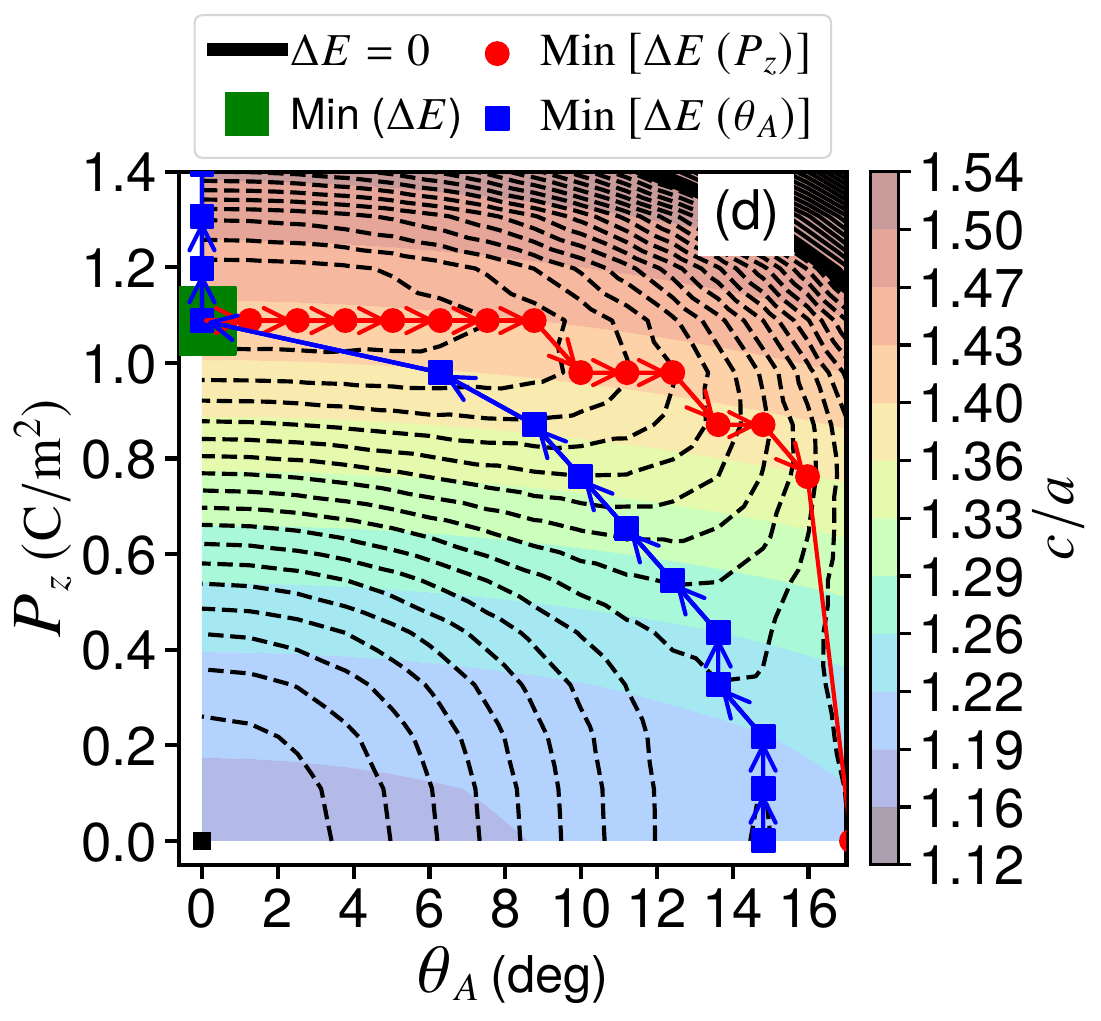}%
    \label{fig:GAcomp8e01}%
  }
  \caption{Energy landscapes for structures obtained by superimposing $A$-point rotations on the extrapolated supertetragonal state (see Eq. \eqref{eq:interpolationdis}) under compressive strains of $-7\%$ [(a), (b)] and $-8\%$ [(c),  (d)].
  Panels (a) and (c) correspond to the undoped system, while panels (b) and (d) correspond to an electron doping level of $0.1~e/\mathrm{f.u.}$. 
  Contours of $\Delta E$ are plotted at $0.04$~eV intervals, with dashed and thin or thick solid lines denoting $\Delta E<0$, $\Delta E<0$ and $\Delta E=0$, respectively.
  The lines with red circles and blue squares mark the path of local minima with fixed A-point or $\Gamma$-point mode, respectively.
  The color map shows the changes in the c/a ratios. 
}
  \label{fig:EGammaA}
\end{figure}

After analyzing the strain and doping coupling of the individual modes separately, we now turn to the interaction between these different soft modes. 
Superimposing $A$ and $M$ mode results in the alternating increase/decrease of the rotation angle. 
This superposition is energetically not favourable, see appendix \cref{fig:EAM}.
Analogously, the modulation of polarization by superimposing $\Gamma$ and $X$ modes is not considered further. 
In the following we thus focus only on the $A$ and $\Gamma$ modes being lowest in energy.

If the $\Gamma$ and $A$ modes would couple cooperatively, the energetic ground state for each value of strain would be given by a linear combination of these. 
Instead, the energy minima of the two-dimensional energy landscape are related to the separate modes, i.e.\ they  either correspond to zero polarization or to zero rotation angle (\cref{fig:EGammaA}). 
As shown by the  color encoding also the  $c/a$ ratio is given by the amplitudes of the separate modes without any collaborative changes. 
On the one hand, the change of $c/a$ with the rotation angle is negligible. 
On the other hand, the polar distortion is coupled to the large increase of $c$ related to the supertetragonal state. 

Depending on strain, the global energy minimum is either the state with $A$-mode rotations (from $-1\%$ to $-7\%$),  while for $\eta \leq -8\%$ it is the polar supertetragonal state.
Note that the energy difference between both modes is only $0.014$~eV at $-7\%$.
While both modes do not couple collaboratively, the (metastable) supertetragonal state can tolerate a small additional $A$-point rotation of about $4^\circ$--$5^\circ$ for $-7\%$ ($\approx6^\circ$ at $-8\%$), while the $A$-rotation state tolerates a polar distortion of roughly $0.2$--$0.3$~$\mathrm{C\,cm^{-2}}$ at both strain values (\cref{fig:EGammaA}(a),(b)).
For larger superimposed rotations, the energy of the system increases abruptly and 
one may expect that one displacement would suppress the other.
However, there is no collaborative coupling.

\cref{fig:EGammaA}[(b),(d)] illustrate the change of the energy landscape by electron doping ($0.1~e/\mathrm{f.u.}$).
As discussed above, this doping lowers the energy for both modes and stabilizes the supertetragonal state relative to the $A$-rotation mode. Therefore, the former is already the ground state for $\eta=-7$~\%.
Once the supertetragonal state is the groundstate it is stable for rotations of up to $-8\%$. 
Nevertheless, the doping does not induce any collaborative coupling between modes.

\section{DISCUSSION AND CONCLUSION}
In this work we revisited the phase stability of \ce{BaSnO3} under compressive biaxial strain by means of density functional perturbation theory simulations.
We have confirmed the transition to a supertetragonal state with large polarization predicted before.
However we have shown that the magnitude of polarization and $c/a$ are not different to  prototypical \ce{BaTiO3} at the same value of strain, instead the main difference is the first order character of the transition from the nonpolar phase at moderate strain to this phase at a large compressive strain.
As soon as this phase is induced it is related to a large energy well and thus a transition is also likely at finite temperatures. 
This may allow to switch the material between a non-polar state and a state with large polarization with small external stimuli.

We have found two possible explainations why the ferroelectric state has not yet been realized in experiment: While the critical strain of $-5\%$ predicted in \cite{Li2025} could still be accessible in superlattices, it has to be interpreted as a lower bound, only.
Changing from GGA to LDA potentials in DFT, increases the magnitude of the critical compressive strain to $-6.2\%$.
More important, we showed that the onset of the ferroelectric state is shifted to more than $-7\%$ as instead a competing A-point instability is realized for moderate values of strain. 

While it is challenging to reach this range of strain, electron doping does not only allow to reduce the critical strain for the onset of a supertetragonal instability, it also stabilizes this state relatitive to octahedral rotation. 
Furthermore, we have shown that an external field may lower the critical strain to stabilize the supertetragonal state and at least at low temperatures this state may be metastable with respect to the competing octahedra rotations if the field is removed.
Upcoming work has to test this hypothesis in finite-temperature simulations. 

\begin{acknowledgments}
B.S. acknowledges financial support from the China Scholarship Council (CSC).

\end{acknowledgments}

\FloatBarrier

\appendix
\section{Equilibrium lattice constants}
\begin{table}[tbp]
  \caption{\label{tab:bulk}%
    Equilibrium lattice constants obtained using different exchange--correlation functionals.
    The first and second rows show calculated and reference values respectively.
    }
  \begin{ruledtabular}
    \begin{tabular}{lllll}
      & \textrm{LDA} & \textrm{PBEsol}  & \textrm{PBE} & \textrm{Expt.} \\
      \colrule
      \multirow{2}{*}{$a_0$(\AA)} 
          & 4.088 & 4.152 & 4.180 & \multirow{2}{*}{4.116 \cite{maekawa2006}} \\
        & 4.070\cite{Monserrat2018}  & 4.139 \cite{Zhang2017} & 4.186 \cite{bevillon2008} & \\[2pt]
    \end{tabular}
  \end{ruledtabular}
\end{table}
To validate our computational setup, in \cref{tab:bulk}, we calculated the equilibrium lattice constant using LDA, PBEsol, and PBE and compared the results with previously reported theoretical and experimental values, showing good overall agreement.
\section{Strain dependence of the \texorpdfstring{$\Gamma$}{Gamma} mode}
\begin{table}[tbp]
  \centering
  \caption{Frequency of the $\Gamma$ mode as a function of strain.}
  \label{tab:strain_frequency}
  \begin{ruledtabular}
    \begin{tabular}{ccccccccc}
      Strain $\eta$ (\%) 
      & -6.7 & -6.6 & -6.5 & -6.4 & -6.3 & -6.2 & -6.1 & -6.0 \\
      \colrule
      $\omega$ (cm$^{-1}$) 
      & -15.5 & 33.8 & 50.0 & 62.0 & 71.9 & 80.5 & 88.1 & 94.3 \\
    \end{tabular}
  \end{ruledtabular}
\end{table}
A denser strain sampling of $0.1\%$ was performed for the polar $\Gamma$-point mode to more accurately determine the onset of the instability, with the corresponding frequencies listed in \cref{tab:strain_frequency}.
\section{Coupling between the \texorpdfstring{$M$}{M}- and \texorpdfstring{$A$}{A}-point rotational modes}
\begin{figure}[tbp]
  \centering
  \subfloat{%
    \includegraphics[width=0.75\linewidth,clip, trim=0cm 2.2cm 0cm 0cm]{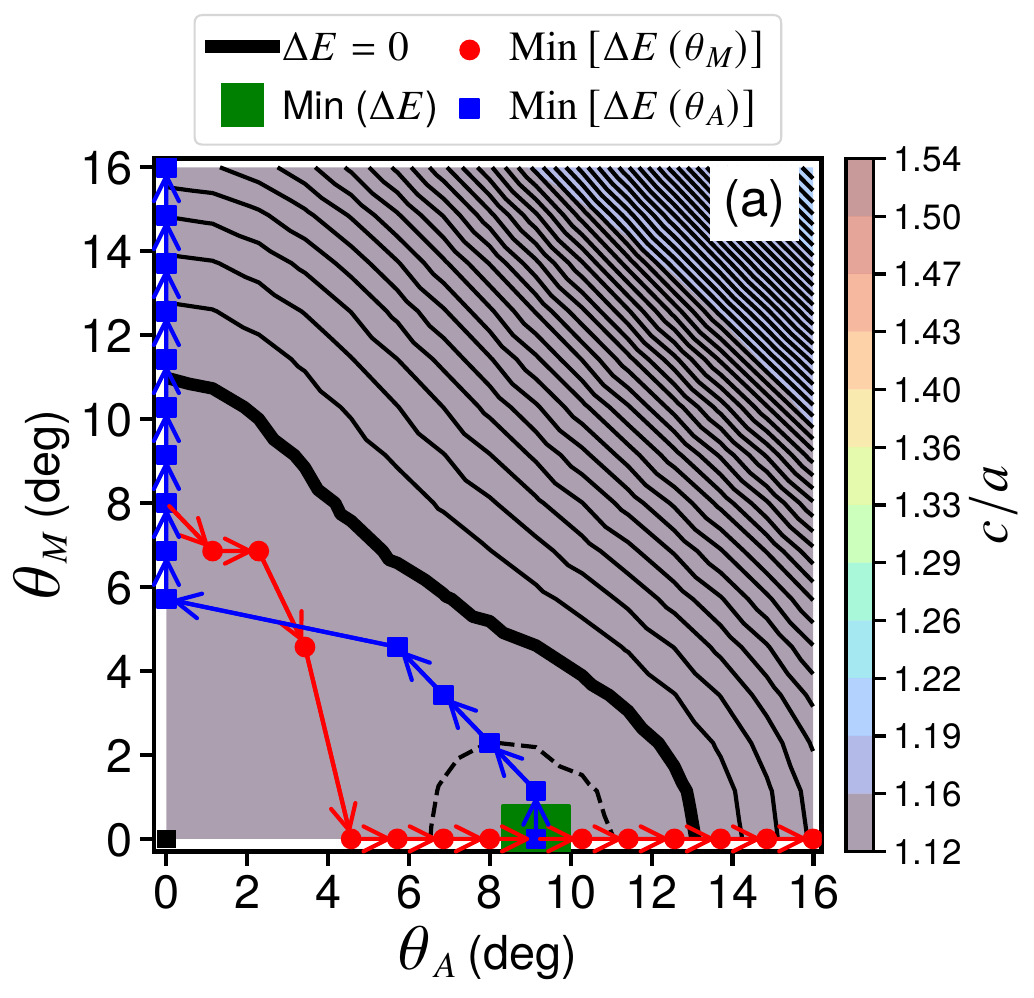}%
    \label{fig:AMcomp5}%
    
  }\vspace{-1em}
  
  \subfloat{%
    \includegraphics[width=0.75\linewidth,clip, trim=0cm 0cm 0cm 2.5cm]{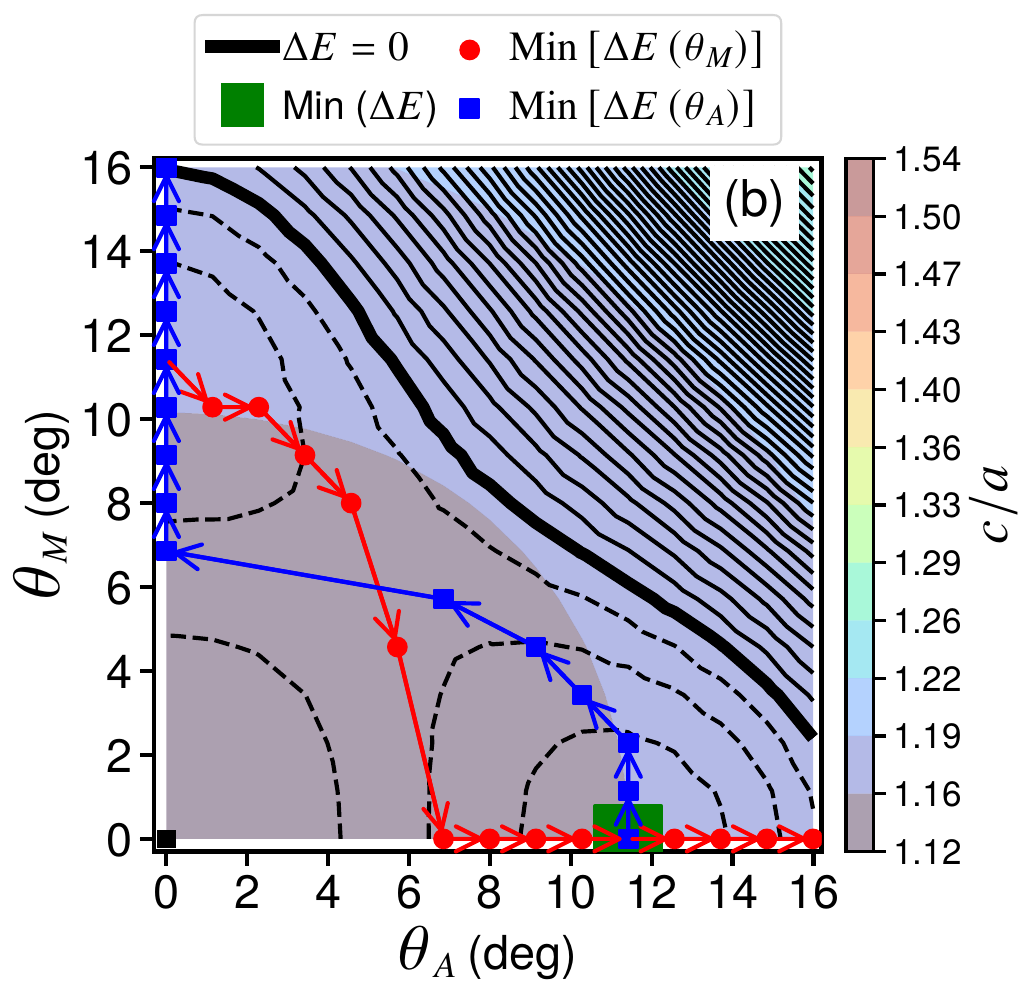}%
    \label{fig:AMcomp7}%
    
  }
  \caption{Energy landscape as a function of the combined M- and A-point mode eigendisplacements under compressive strains of $-5\%$(a), $-7\%$(b). 
  Contours are plotted at 0.04 eV intervals (thin lines), with dashed and solid lines denoting negative and positive values, respectively.
  The thick black line marks the energy of the high-symmetry state ($\Delta E=0$).
  The lines with red circles and blue squares mark the path of local minima with fixed A- or M-mode, respectively.
   The color map shows the changes in the $c/a$-ratios.
  }
  \label{fig:EAM}
\end{figure}
The coupling between the $M$- and $A$-point rotational modes was further examined through two-dimensional energy landscapes at representative compressive strains, as shown in \cref{fig:EAM}.
\section{Electronic density of states}
\begin{figure}[tbp]
  \centering
  \includegraphics[width=0.65\linewidth]{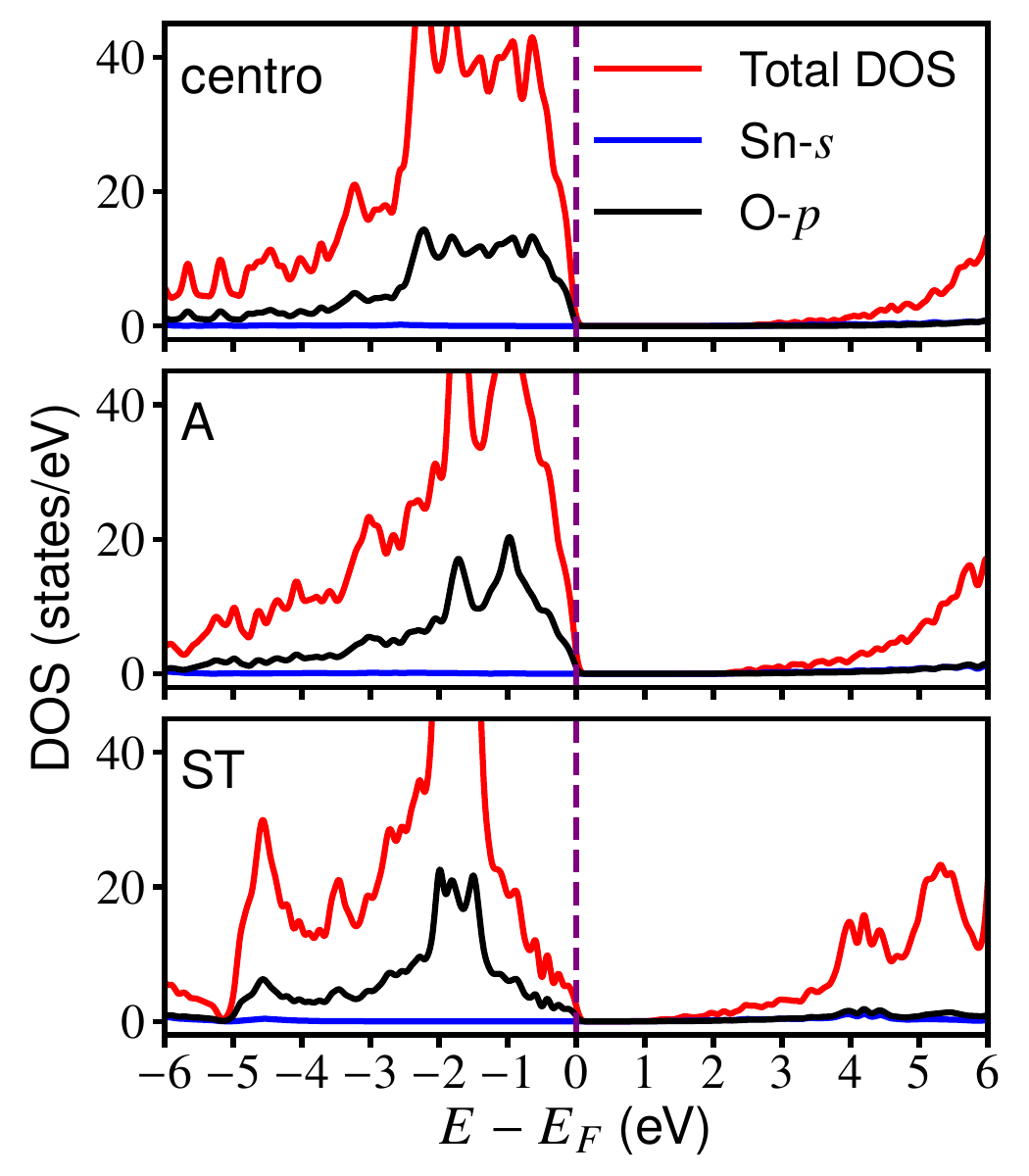}
  \caption{Density of states of the centrosymmetric tetragonal (centro), $A$-derived rotational, and supertetragonal (ST) structures of BSO at $-8\%$ compressive strain.
  The total DOS (red) and the Sn-$s$ (blue) and O-$p$ (black) projected DOS are shown, with energies referenced to the respective Fermi levels. 
  Since the $M$- and $A$-derived rotational states show very similar electronic structures, only the $A$-derived state is shown here. 
  } 
  \label{fig:dos}
\end{figure}
The electronic structures of the centrosymmetric, rotational, and supertetragonal states were compared through their densities of states, as shown in \cref{fig:dos}.

\clearpage

\bibliographystyle{apsrev4-2}
\bibliography{References}

\end{document}